\documentstyle[12pt,epsf]{article}
\newcommand{\vs}{\vspace{-0.25cm}}
\newcommand{\beq}{\begin{equation}}
\newcommand{\eeq}{\end{equation}}
\newcommand{\beqa}{\begin{eqnarray}}
\newcommand{\eeqa}{\end{eqnarray}}

\begin{document}
\hfill{FZJ--IKP--TH--2001--23}
\bigskip
\begin{center}
{{\Large\bf Complete next-to-leading order calculation for pion production
in nucleon-nucleon collisions at threshold}}
\end{center}
\bigskip
\begin{center}
{\large C. Hanhart $^a$ and N. Kaiser $^b$}

\medskip

{\it a: Forschungszentrum J\"ulich, Institut f\"ur Kernphysik 
(Theorie)\\ D-52425 J\"ulich, Germany\\}
{\it b: Institut f\"ur Theoretische Physik, Physik--Department T39,\\
Technische Universit\"at M\"unchen, D--85747 Garching, Germany \\}
\end{center}
\bigskip
\begin{abstract}
Based on a counting scheme that explicitly takes into account the large 
momentum $\sqrt{M m_\pi}$  characteristic for pion production in  
nucleon-nucleon collisions we calculate all diagrams for the reaction $NN \to
NN\pi$ at threshold up to next-to-leading order. At this order there
are no free parameters and the size of the next-to-leading order contributions 
is in line with the expectation from power counting. The 
sum of loop corrections at that order vanishes for the process $pp \to pp\pi^0$
at threshold.
The total contribution at next-to-leading order from loop diagrams
that include the delta degree of freedom
vanishes  at threshold in both reaction channels
$pp\to pp\pi^0,\,pn\pi^+$.
\end{abstract}

\vskip 1.cm

The high precision data for the processes $pp \to pp\pi^0$, $pp \to pn\pi^+$ 
and $pp\to d\pi^+$ in the threshold region \cite{DATA} have spurred a flurry 
of theoretical investigations. The first data on neutral pion production were 
a big surprise because the experimental cross sections turned out to be a 
factor of five larger than the theoretical predictions based on direct pion 
production and neutral pion rescattering fixed from on-shell $\pi N$ data 
\cite{misau,niskan}. Subsequently, it was argued that heavy-meson exchanges 
might be able to remove this discrepancy \cite{LR}. On the other hand, it was 
found \cite{oset,unsers} that the (model-dependent) off-shell behavior of the 
full $\pi N$ T-matrix can also enhance the cross sections near threshold 
considerably. 

Due to their nature as pseudo-Goldstone bosons the dynamics of pions is largely
constrained by chiral symmetry. Thus one might hope that effective field theory
studies which incorporate these constraints strictly will help to resolve the 
so far confusing situation. In the literature there are several calculations 
carried out in the framework of tree-level chiral perturbation theory including
the dimension two (single-nucleon) operators for neutral pion production 
\cite{pmmmk,bira1,bira2,lee} as well as for charged pion production 
\cite{HHH,CP4}. A common feature of these calculations is that the 
contributions from the isoscalar pion rescattering interfere destructively with
the direct production amplitude, thus leading to an even more severe 
discrepancy between experiment and theory. It should be noted that such an 
interference pattern is in contradiction to the one found in phenomenological 
approaches \cite{oset,unsers}. Furthermore, within
the Weinberg scheme, where all momenta are considered of
the order of $m_\pi$, one loop calculations have been 
performed for neutral pion production $pp\to pp\pi^0$ 
\cite{CPloop1,CPloop2,ando}. According to some of these works the loop 
corrections are larger by at least a factor of two compared to the tree level 
diagrams,
that according to the counting scheme applied appear one order down.
 This feature (if correct) would seriously question the convergence of
the chiral expansion for pion production in NN-collisions. On the other 
hand, according to ref.\cite{ourpwave} the chiral expansion seems to show
convergence in the case of p-wave pion production.

The purpose of the present work is to present a complete next-to-leading order
calculation of the reaction $NN\to NN\pi$ at threshold. In particular, we
evaluate all one-loop diagrams at next-to-leading order employing a counting 
scheme that takes into account the large momentum $\sqrt{Mm_\pi}$ 
characteristic for pion production in NN-collisions, as
suggested in refs. \cite{bira1,ourpwave}. We consider also the
contributions from explicit delta-isobars at tree level and at one-loop order. 
To the order we are working there are no free parameters and we demonstrate
that the size of the individual next-to-leading order contributions is in line with the
expectations from power counting. 

Let us begin with writing down the general form of the threshold T-matrix for
the pion production reaction $N_1(\vec p\,)+N_2(-\vec p\,) \to N+N +\pi$ in 
the center-of-mass frame, which reads \cite{novel}: 
\begin{eqnarray}
T^{cm}_{th}(NN\to NN \pi) &=& {{\cal A}\over 2} \, ( i\,\vec
\sigma_1 - i\, \vec \sigma_2+\vec \sigma_1 \times \vec \sigma_2)\cdot \vec p
\,\,\, (\vec \tau_1+\vec \tau_2)\cdot \vec \phi^{\,*} \nonumber \\ & & +
{{\cal B} \over 2} \,(\vec \sigma_1 + \vec \sigma_2)\cdot \vec p  \,\,\, (i\,
\vec \tau_1- i\,\vec \tau_2 + \vec \tau_1 \times \vec \tau_2)\cdot \vec
\phi^{\,*} \,,  \end{eqnarray}
with $\vec\sigma_{1,2}$ and $\vec\tau_{1,2}$ the spin and isospin operators of
the two nucleons. $\vec\phi$ denotes the three-component isospin wave function
of the final state pion produced in an s-wave state, e.g. $\vec \phi= (0,0,1)$
for $\pi^0$-production and $\vec \phi = (1,i,0)/\sqrt2$ for $\pi^+$-production.
The complex amplitudes ${\cal A}$ and ${\cal B}$ belong to the transitions 
$^3P_0 \to\,^1S_0$ and $^3P_1 \to \,^3S_1$ in the two-nucleon system, 
respectively. In fact the selection rules which follow from the conservation of
parity, angular momentum and isospin allow only for these two transitions for
the reaction $NN\to NN\pi$ at threshold. In the case of neutral pion production
$pp\to pp\pi^0$ the threshold amplitude ${\cal A}$ is the only relevant one 
whereas in charged pion production $pp \to pn \pi^+$ both threshold amplitudes 
${\cal A}$ and ${\cal B}$ can contribute. Note that the threshold T-matrix
written in eq.(1) incorporates the Pauli exclusion principle since combined 
left multiplication with the spin-exchange operator $(1+\vec\sigma_1\cdot\vec 
\sigma_2)/2$ and the isospin exchange operator $(1+\vec\tau_1\cdot\vec\tau_2)
/2$ reproduces $T^{cm}_{th}(NN\to NN \pi)$ up to an important minus sign. The 
magnitude of the nucleon center-of-mass momentum $\vec p$ necessary to produce
a pion at rest is given by: 
\begin{equation} |\vec p\,| = \sqrt{m_\pi(M+m_\pi/4)}\,,\end{equation}
with $M=939\,$MeV and $m_\pi=139.6\,$MeV denoting the nucleon and pion mass, 
respectively. Eq.(2) exhibits the important feature of the reaction $NN\to 
NN\pi$, namely the large momentum mismatch between the initial and the final 
nucleon-nucleon state. This leads to a large invariant (squared) momentum 
transfer $t=-Mm_\pi$ between in- and outgoing nucleons. The appearance of the 
large momentum scale $\sqrt{Mm_\pi}$ in pion production demands for a change 
in the chiral power counting rules, as pointed out already in ref.\cite{bira1}.
In addition, it seems compulsory to include the delta-isobar as an explicit
degree of freedom, since the delta-nucleon mass difference $\Delta = 293\,$MeV
is comparable to the external momentum $p\simeq \sqrt{M m_\pi} =362\, $MeV. The
hierarchy of scales
\begin{equation}
M \gg p \simeq \Delta \gg m_\pi \ ,
\label{schi}
\end{equation}
 suggested by this feature is
in line with findings within meson exchange models where the delta-isobar gives
significant contributions even close to the threshold \cite{jounidel,ourdelta}.

Let us now state our counting rules. The external momentum $p\simeq \sqrt{M
m_\pi}$ sets the overall scale relevant for the process $NN\to NN\pi$. This
momentum scale $p$ enters the internal lines of tree and loop diagrams. 
Therefore we count all four-momenta\footnote{Baryon energies are residual 
energies with the nucleon mass $M$ subtracted.} $l_\mu$ inside loops 
generically as order $p$ and the loop integration measure $\int d^4l$ as order 
$p^4$. A pion propagator is counted as order $1/p^2$. The delta-propagator of 
the form $1/(energy-\Delta)$ counts as order $1/p$, since we made the choice 
$\Delta \sim p$. For the nucleon propagator of the form $1/energy$ one has to 
distinguish whether it occurs outside or inside a loop. The associated residual
energy counts as order $m_\pi$ outside a loop and as order $p \sim 
\sqrt{Mm_\pi}$ inside a loop. Furthermore, external pion energies are counted 
as order $m_\pi$.      

According to these counting rules one-loop diagrams contribute at order $p^2$ 
in the expansion of the T-matrix and thus generate threshold amplitudes of the
form ${\cal A}, {\cal B} \sim p \simeq \sqrt{M m_\pi}$. The new counting rules 
demand also for a reordering of the terms in the interaction Lagrangian, since
"relativistic corrections" proportional to nucleon kinetic energies $p^2/M$ are
now of the same order as "leading order contributions" proportional to residual
nucleon energies. Several examples of this effect will be encountered here. 

In Fig.\,1, we display tree-level diagrams which according to the
abovementioned counting rules contribute at leading order, next-to-leading
order and next-to-next-to-leading order. Diagrams for which the role of both
nucleons is interchanged and diagrams with crossed outgoing nucleon lines are
not shown. Subsets of four diagrams obtained by these operations map properly
onto the crossing antisymmetric threshold T-matrix eq.(1). Diagram a) involving
the (isovector) Weinberg-Tomozawa $\pi\pi NN$-contact vertex gives a
leading-order contribution of the form:
\begin{equation} {\cal A}^{(WT)} = 0\,, \qquad {\cal B}^{(WT)} = -{g_A \over 2M
f_\pi^3} \,, \end{equation}
with $g_A\simeq 1.3$ the nucleon axial vector coupling and $f_\pi = 92.4\,$MeV
the pion decay constant. It is important to note that the Weinberg-Tomozawa
vertex generates here a proportionality factor $m_\pi$ at "leading order" in
the chiral $\pi N$-Lagrangian via the pion and nucleon (residual) energies as
well as through a "relativistic correction" of the form $p^2/M$. This factor of
$m_\pi$ gets finally canceled by the pion propagator $[m_\pi(M+m_\pi)]^{-1}$. 
Obviously, the isovector Weinberg-Tomozawa vertex cannot contribute to the
neutral pion production threshold amplitude ${\cal A}$. From the one-pion
exchange diagram b) one finds:
\begin{equation} {\cal A}^{(1\pi)}={g_A^3 \over 8Mf_\pi^3}  \,, \qquad 
{\cal B}^{(1\pi)} = {3g_A^3 \over 8M f_\pi^3} \,. \end{equation}
This result stems from the recoil correction to the $\pi NN$-vertex
proportional to $(m_\pi/M)\,\vec \sigma_1 \cdot \vec p$ with the $m_\pi$-factor
getting now canceled by the intermediate nucleon propagator. Furthermore, the 
product of the two vertices on the left nucleon line $(\vec \sigma_1 \cdot \vec
p\,)^2 = Mm_\pi$ is canceled by the pion propagator. The ratio ${\cal B}^{(1
\pi)}/{\cal A}^{(1\pi)}=3$ has its origin in the isospin factor of diagram b).
From the analogous diagram d) with one virtual delta-isobar excitation one 
finds: 
\begin{equation} {\cal A}^{(\Delta)} = {g_A^3 m_\pi \over 4M f_\pi^3 \Delta} 
\,, \qquad {\cal B}^{(\Delta)} = 0\,, \end{equation}
where we have used the empirically well satisfied relation $h_A=3g_A/\sqrt{2}$
for the $\pi N\Delta$-coupling constant. The spin and isospin transition 
operators entering the $\pi N\Delta$-vertex $(h_A/2f_\pi)\vec S\cdot\vec p\,
T_a$ satisfy the usual relations $S_i S_j^\dagger=(2\delta_{ij}-i\epsilon_{ijk}
\sigma_k)/3$ and $T_a T_b^\dagger=(2\delta_{ab}-i\epsilon_{abc}\tau_c)/3$. The
latter isospin relation is the reason behind the vanishing of ${\cal B}^{(
\Delta)}$. According to our counting of the mass-splitting $\Delta$ the term 
${\cal A}^{(\Delta)}$ in eq.(6) is a next-to-leading order contribution, since
$\Delta \sim p$ (c.f. relation (\ref{schi})). Diagram f) involves the second 
order chiral $\pi\pi
NN$-contact vertex proportional to the low-energy constants $c_{1,2,3,4}$
\cite{bkmnun}. We find the following contributions to the threshold amplitudes
at next-to-next-to-leading order:
\begin{equation} {\cal A}^{(c_i)} = {g_Am_\pi \over 2M f_\pi^3}(c_3+2c_2-4c_1)
\,, \qquad  {\cal B}^{(c_i)} = {g_A m_\pi \over 2M f_\pi^3} (c_4+c_3+2c_2-4
c_1)\,. \end{equation}
In a previous calculation in ref.\cite{pmmmk} (see eq.(32) therein) the 
$c_2$-term has been found with a relative factor 1/2 smaller. The reason for
this discrepancy is again that "relativistic corrections" from the $c_2$-vertex
are of the same order as its "static" contribution, since $p^2/M = m_\pi$. We
also note that our results eqs.(4-7) agree up to the respective order with
those of the fully relativistic calculation in ref.\cite{novel} where no
approximations to the threshold kinematics have been made. We do not specify
here the contributions from diagrams c), e) and g) in Fig.\,1 which are
proportional to the (a priori unknown) strengths of four-nucleon
contact-vertices etc. It is important to note that already at leading order 
long-range effects from pion-exchange and short-range contributions appear 
simultaneously. 

Let us now turn to the non-vanishing one-loop diagrams at threshold. Not every
loop diagram appearing formally at next-to-leading order truly contributes at 
that order. In case of the diagrams b) and c) in Fig.\,2 the (spin-independent)
one-loop $\pi N$-scattering subdiagrams are proportional to $m_\pi^3$, and this
pushes their contributions to the threshold T-matrix eq.(1) beyond
next-to-leading order. A closer inspection of diagrams a) in Fig.\,2 reveals 
that they contribute in the form $m_\pi \ln m_\pi$ to the threshold amplitudes
${\cal A}$ and ${\cal B}$, i.e. beyond next-to-leading order. The specific 
vertex structures of diagrams d) and e) in Fig.\,2 make also their 
next-to-leading order contributions vanishing. Therefore we have to focus only
on the diagrams shown in Figs.\,3 and 4.

We evaluate only the genuine next-to-leading order pieces of the loop integrals
emerging from the diagrams in Figs.\,3 and 4. For instance, in the 
integrands we
can systematically drop terms of order $m_\pi$ compared to $l_0$ (and
$\Delta$). Straightforward but tedious evaluation leads to the following 
next-to-leading order contributions of the one-loop diagrams in Fig.\,3 with
nucleons only:   
\begin{equation} {\cal A}^{(N-loop)} = {g_A^3 \sqrt{Mm_\pi} \over 256 f_\pi^5}
(-2-1+3)\,, \qquad  {\cal B}^{(N-loop)} = {g_A^3 \sqrt{Mm_\pi} \over 256 
f_\pi^5}(-2+0+3)\,. 
\label{nucres} \end{equation}
Here, the numerical entries correspond to the diagrams a), b) and c), in that
order. Interestingly, the total next-to-leading order loop contribution vanishes
identically for neutral pion production ${\cal A}^{(N-loop)}=0$.
Diagrams a) and c) in Fig.\,3 have been calculated fully
relativistically (i.e. without any approximation to the threshold kinematics) 
for $pp\to pp\pi^0$ in ref.\cite{novel}. It is an important check for our 
calculation that the non-analytical piece proportional to $\sqrt{Mm_\pi}$ 
agrees with the one derived by expanding eq.(16) in ref.\cite{novel}. 
In addition, after
correcting a sign error in ref.\cite{CPloop2} and extracting at threshold the
truly next-to-leading order pieces from that work our results agree with 
theirs \cite{fredpriv}. 

Numerically, the loop correction in eq.(\ref{nucres}) gives 
${\cal B}^{(N-loop)} = g_A^3 
\sqrt{Mm_\pi}/256 f_\pi^5 \simeq 0.70\,{\rm fm}^4$. This is about 50\% of the
leading order one-pion exchange contributions $|{\cal B}^{(WT)}|\simeq
1.33\,{\rm  fm}^4$ or ${\cal B}^{(1\pi)}\simeq 1.69\,{\rm fm}^4$. Indeed from 
chiral power counting one expects a similar suppression factor $p/M= 
\sqrt{m_\pi/M} \simeq 0.4$.  

According to our counting of the mass difference $\Delta \sim \sqrt{Mm_\pi}$
loop diagrams with explicit delta-isobars are of the same order as those with
nucleons only, namely of order $p^2$. The relevant one-loop diagrams which 
generate truly next-to-leading order contributions are shown in Fig.\,4. 
Straightforward but tedious evaluation leads to the following result:    
\begin{equation} {\cal A}^{(\Delta-loop)} = {g_A^3 K(\Delta)\over 32 f_\pi^5}
(8-12+1+3) \,,\qquad {\cal B}^{(\Delta-loop)} = {g_A^3  K(\Delta)\over 32 
f_\pi^5} (8-12+3+1) \end{equation}
with the numerical entries corresponding to the subclasses a), b), c) and d), 
in that order. The relevant combination of loop functions reads:
\begin{equation} K(\Delta)= 2J_0(-\Delta) -2\Delta\, I_0(-Mm_\pi) +(2\Delta^2-
Mm_\pi)\,\gamma_0(-\Delta,-Mm_\pi) \,,\end{equation}
with the following loop integrals \cite{bkmnun} truncated at lowest order 
according to our counting scheme:
\begin{equation} J_0(-\Delta) = 4\Delta \, L(\lambda) +{\Delta \over 4\pi^2}
\bigg( \ln{2 \Delta \over \lambda} -{1\over 2} \bigg) \sim {\cal O}(p)\,, 
\end{equation}
\begin{equation}I_0(-Mm_\pi) =-2 L(\lambda) -{1\over 16 \pi^2} \bigg( 1+ \ln{ 
Mm_\pi \over \lambda^2} \bigg) \sim {\cal O}(p^0)\,, \end{equation}
\begin{equation}\gamma_0(-\Delta,-Mm_\pi) = {1\over 4\pi^2 \sqrt{Mm_\pi}}
\int_0^\infty {dx \over 1+x^2} \arctan{x \sqrt{Mm_\pi}  \over 2 \Delta } \sim
{\cal O}(p^{-1})\,. \end{equation}
The (scale dependent) quantity:  
\begin{equation} L(\lambda) = {\lambda^{d-4} \over 16\pi^2} \bigg[ {1\over d-4}
+{1\over 2} (\gamma_E-1-\ln 4\pi) \bigg] \,, \end{equation}
denotes for the standard divergent piece in dimensional regularization. The
reason for grouping together the three specific diagrams into subclass b) is 
that this way the (in the chiral limit) singular term $\Delta^2 J_0(-\Delta)/
Mm_\pi$ does not appear explicitly. Evidently, the formal limit $\Delta \to 0$
corresponds to loops diagrams with nucleons only, and therefore $K(0)=
-\sqrt{Mm_\pi}/16$ enters eq.(8). Note, however, that for planar box diagrams 
this limit becomes inconsistent with the counting scheme employed.

One concludes that the contributions stemming from loop diagrams with 
delta-excitation vanish identically for both threshold amplitudes ${\cal A}$ 
and ${\cal B}$, respectively for both reaction channels $pp \to pp\pi^0, pn 
\pi^+$. The complete cancellations in eq.(9) are actually important consistency 
checks for our power counting scheme $\Delta \sim p$. The combination of loop 
functions $K(\Delta) \sim p$ in eq.(10) is divergent, but at next-to-leading 
order there is no local counter term to absorb divergences.

In summary, we have performed here a complete next-to-leading order calculation
of the reaction $NN\to NN\pi$ at threshold. We have employed the counting
scheme developed in refs. \cite{bira1,ourpwave},
that explicitly accounts for the large momentum $p \simeq \sqrt{Mm_\pi}$
characteristic for this process. We find that the total next-to-leading order 
loop corrections either vanish or are in accordance with the
expectation from power counting.
At this stage we conclude, that the chiral expansion seems to converge
also in the s-wave. Note, however, that  at next-to-next-to-leading order a 
large number of loops enters, that have not yet been evaluated completely.

In order to compare our results directly to
pion production data the emerging chiral operators have to be folded with 
(realistic) NN-wave functions. This convolution has been carried out in 
ref.\cite{ando} in a way that the symmetries are preserved. However, in that 
work the traditional Weinberg counting has been used. Consequently the results
presented in ref.\cite{ando} do not allow any firm conclusion about the 
convergence of the chiral series, since contributions of different orders are 
mixed and the next-to-next-to-leading order is incomplete. Based on our 
counting scheme a complete next-to-next-to-leading order calculation is within
reach and should be performed. Another direction should be the calculation of
loop corrections to the higher partial waves amplitudes.

\bigskip

\begin{figure}
\vspace{10.5cm}
\includegraphics{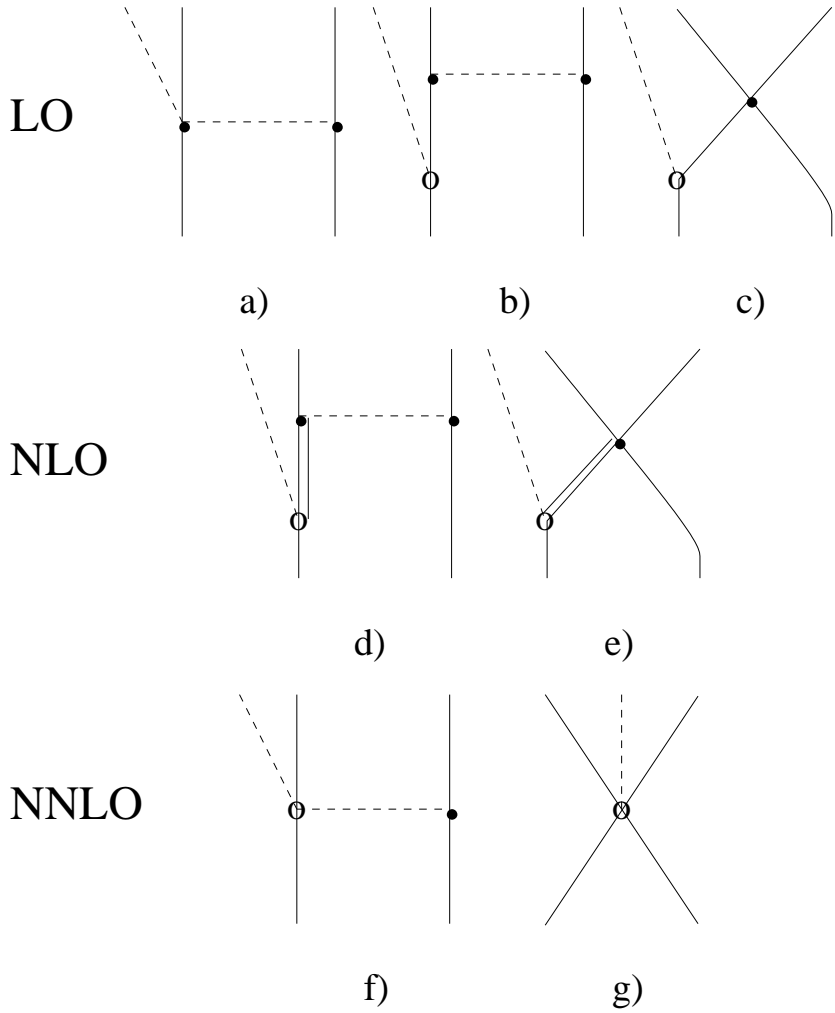}
\caption{Tree level contributions to threshold pion production at leading order
(a,b,c), next-to-leading order (d,e) and next-to-next-to-leading order (f,g). 
A single solid, double solid and dashed line denotes a nucleon, delta-isobar
and pion, respectively. Leading (subleading) order vertices are symbolized by 
solid dots (open circles).}
\label{LO} 
\end{figure}

\bigskip

\begin{figure}
\vspace{12.cm}
\includegraphics{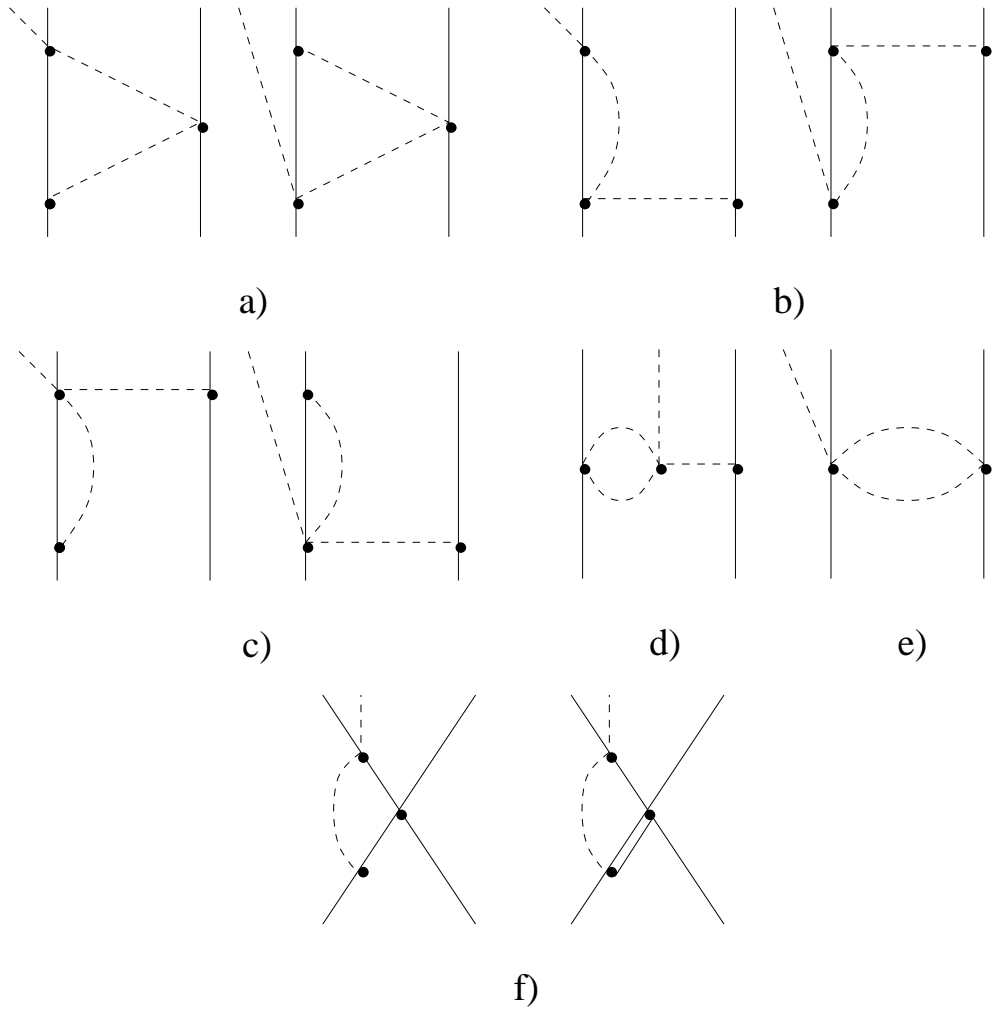}
\caption{One-loop diagrams that start to contribute at next-to-next-to-leading
order. For further notations see Fig.\,1.}
\label{NNLO} 
\end{figure}

\bigskip

\begin{figure}
\vspace{4.5cm}
\includegraphics{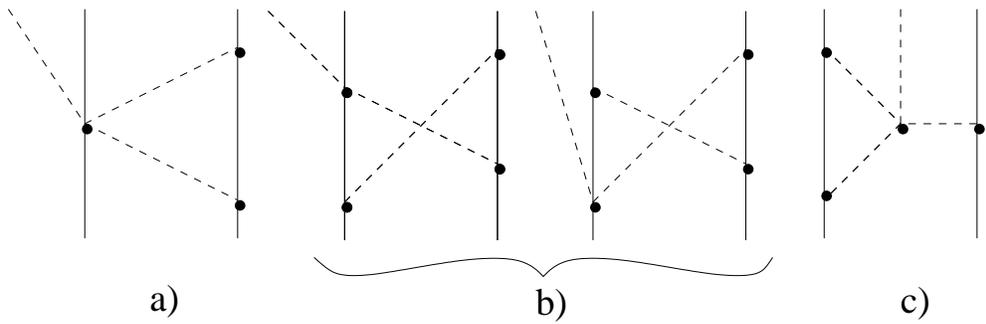}
\caption{Next-to-leading order one-loop diagrams for pion production at 
threshold with nucleons only. For further notations see Fig.\,1.}
\label{NLO} 
\end{figure}

\bigskip

\begin{figure}
\vspace{9cm}
\includegraphics{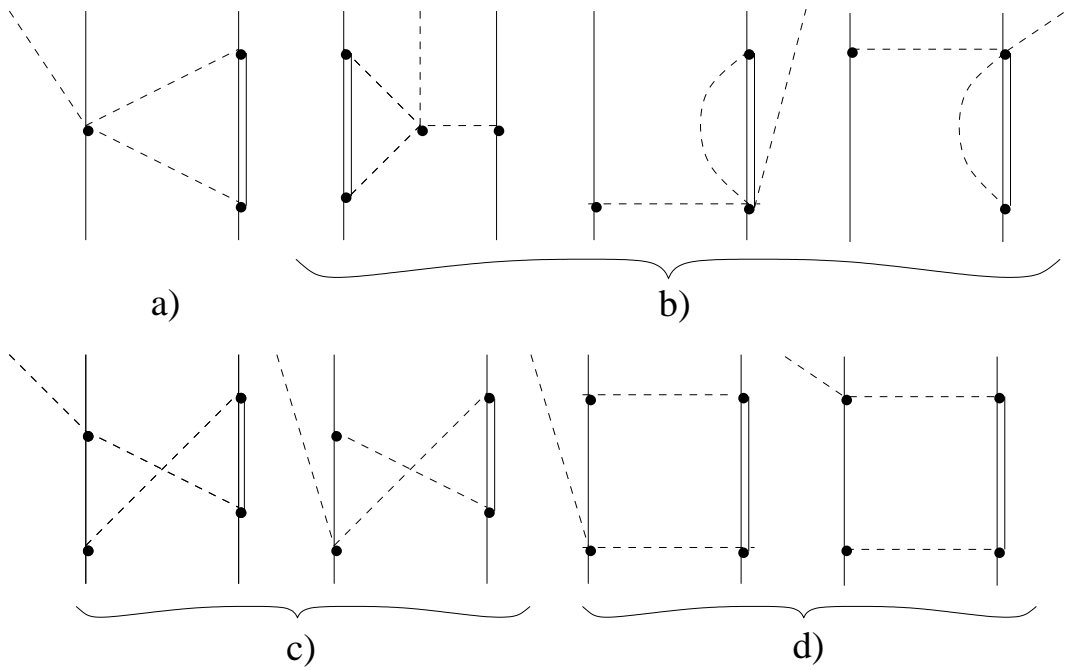}
\caption{Next-to-leading order one-loop diagrams for pion production at
threshold with intermediate delta-isobars. For further notations see Fig.\,1.}
\label{NLO_del} 
\end{figure}
\end{document}